\def \SAIT #1 #2 {{\em Mem.\ Soc.\ Astron.\ It.\/} {\bf #1}, #2}
\def \MESS #1 #2 {{\em The Messenger\/} {\bf #1}, #2}
\def \ASTRNACH #1 #2 {{\em Astron. Nach.\/} {\bf #1}, #2}
\def \AAP #1 #2 {{\em Astron. Astrophys.\/} {\bf #1}, #2}
\def \AAL #1 #2 {{\em Astron. Astrophys. Lett.\/} {\bf #1}, L#2}
\def \AAR #1 #2 {{\em Astron. Astrophys. Rev.\/} {\bf #1}, #2}
\def \AAS #1 #2 {{\em Astron. Astrophys. Suppl. Ser.\/} {\bf #1}, #2}
\def \AJ #1 #2 {{\em Astron. J.\/} {\bf #1}, #2}
\def \ANNREV #1 #2 {{\em Ann. Rev. Astron. Astrophys.\/} {\bf #1}, #2}
\def \APJ #1 #2 {{\em Astrophys. J.\/} {\bf #1}, #2}
\def \APJL #1 #2 {{\em Astrophys. J. Lett.\/} {\bf #1}, L#2}
\def \APJS #1 #2 {{\em Astrophys. J. Suppl.\/} {\bf #1}, #2}
\def \APSS #1 #2 {{\em Astrophys. Space Sci.\/} {\bf #1}, #2}
\def \ASR #1 #2 {{\em Adv. Space Res.\/} {\bf #1}, #2}
\def \BAIC #1 #2 {{\em Bull. Astron. Inst. Czechosl.\/} {\bf #1}, #2}
\def \JSQRT #1 #2 {{\em J. Quant. Spectrosc. Radiat. Transfer\/} {\bf #1}, #2}
\def \MN #1 #2 {{\em Mon. Not. R. Astr. Soc.\/} {\bf #1}, #2}
\def \MEM #1 #2 {{\em Mem. R. Astr. Soc.\/} {\bf #1}, #2}
\def \PLR #1 #2 {{\em Phys. Lett. Rev.\/} {\bf #1}, #2}
\def \PASJ #1 #2 {{\em Publ. Astron. Soc. Japan\/} {\bf #1}, #2}
\def \PASP #1 #2 {{\em Publ. Astr. Soc. Pacific\/} {\bf #1}, #2}
\def \NAT #1 #2 {{\em Nature\/} {\bf #1}, #2}

\documentstyle{memsait}
\input epsf.sty

\def\BmV0{\mbox{(B-V)$^{\rm o}$}}
\def\VmK0{\mbox{(V-K)$^{\rm o}$}}
\def\MV0{\mbox{M$_{\rm V}^{\rm o}$}}

\def\carbiso{\mbox{${\rm ^{12}C/^{13}C}$}}
\def\etal{\mbox{{\it et al.}}}
\def\eg{\mbox{{\it e.g.}}}

\begin{opening}
\title{DO AGB STARS DIFFER CHEMICALLY FROM RGB STARS IN GLOBULAR CLUSTERS?} 
\author{CHRISTOPHER SNEDEN$^1$, INESE I. IVANS$^1$, ROBERT P. KRAFT$^2$}
\institute{$^1$Department of Astronomy and McDonald Observatory,
University of Texas, Austin, TX 78712, USA\\
$^2$UCO/Lick Observatory, Board of Studies in Astronomy and
Astrophysics, University of California, Santa Cruz, CA  95064, USA}
\date{} 
\end{opening}

\begin{document}

\oddpagefooter{}{}{} 
\evenpagefooter{}{}{} 
\ 
\bigskip

\begin{abstract}

The recent improvements in globular cluster colour-magnitude diagrams, 
coupled with an increase in large-sample spectroscopic abundance 
studies of cluster giants, finally allow some attempts at a
statistically meaningful comparison of the chemical compositions
of red giant branch and asymptotic branch cluster stars.  
We review some of the extant data here, concluding that in a few 
clusters the AGB stars show on average smaller amounts of 
high-temperature proton-capture synthesis products (low oxygen, 
high sodium and aluminum) at their surfaces than do the 
first-ascent RGB stars.  
This suggests that those RGB stars with envelopes that have been
enriched with proton-capture material also have high helium contents.  
Such stars after the He-flash then take up
residence on the bluest parts of the HB (as a consequence of their
high envelope helium), probably never to return to the AGB during
subsequent evolutionary stages.

\end{abstract}

\section{Introduction}
Asymptotic giant branch (AGB) stars of all populations have basically
the same interior structures, with shell fusion zones of He and H 
surrounding C-O cores.
Most AGB stars will undergo episodes of mass loss that eject their outer
envelopes, leaving the exposed cores to fade away as white dwarfs.
Thus low mass, metal-poor halo AGB stars and their higher mass,
metal-rich disk counterparts exist in the same evolutionary domains
and share the same eventual fates; all these stars appear
to be theoretically quite similar.
 
Observationally however, the properties of disk and halo population 
AGB stars are quite distinct.
The high mass, high metallicity AGB stars are both extremely luminous and
extremely cool.
Sometimes they are surrounded by substantial gas/dust shells of
their own making, and thus present unique photometric signatures (especially
in the infrared).
Often they exhibit spectroscopic peculiarities (strong carbon-containing
molecular and neutron-capture element features) indicative
of nuclear processing in their He fusion zones.
These are the stars that are treated in most of the contributions to this
workshop.

In contrast, the low mass, low metallicity stars that can be positively
associated with the AGB are photometrically and spectroscopically somewhat
difficult to distinguish from first-ascent red giant branch (RGB) stars.
In globular cluster colour-magnitude (c-m) diagrams the AGB is a 
thinly-populated stream of stars connecting the red end of the 
horizontal branch (HB) and the end of the RGB.
Over the brightest 1--2 magnitudes, the AGB and RGB are separated 
by less than 0.2 in V magnitude at a given B--V colour
(a particularly clear example is the M3 c-m
diagram of Buonanno \etal\ 1986).
Globular cluster AGB stars will not become extremely luminous because they
are former HB stars, whose masses cannot in theory exceed
$\mathcal{M}$~$\sim$~0.6$\mathcal{M}$$_{\odot}$;
the second-ascent AGB tip effectively merges with the first-ascent RGB tip.
Unfortunately, for many globular clusters, photometry precise enough
to cleanly separate the AGB from the RGB for most candidate stars
still does not exist.
The spectra of globular cluster AGB stars also do not differ 
radically from those of RGB stars.
CH stars, those possessing spectroscopic evidence of having possibly 
mixed He shell burning products (carbon, neutron-capture elements) to 
their surfaces, are apparently very rare in globular clusters;
only a handful have been discovered (\eg, McClure \& Norris 1977; 
Cowley \& Crampton 1985; Vanture \& Wallerstein 1992; C\^ote \etal\ 1997).
 
In recent years photometrists and spectroscopists have combined efforts
to substantially increase the quantity and quality of data on AGB stars
in globular clusters.
In this paper, we look for chemical composition differences between
AGB and RGB stars in three globular clusters, concluding that
there is some evidence suggesting that AGB stars have less
chemically evolved surface layers.
This suggestion is then related to the ``second parameter problem''
of globular clusters.

\section{Inter- and Intra-Cluster Chemical Inhomogeneities: A Brief Sketch}

Several decades of spectroscopic investigations have established the
reality of large-scale star-to-star abundance variations among light
elements in globular cluster stars.
The variations are not of the same magnitude in all clusters, and
indeed each cluster seems to have a chemical composition signature that
is not repeated exactly in other clusters.
Most of the abundance inhomogeneities observed in globular clusters 
involve some aspects of so-called ``proton-capture'' nucleosynthesis.
Extensive reviews of these abundance variations have been published
by \eg, Kraft (1979), Freeman \& Norris (1981), Smith (1987),
Suntzeff (1993), Briley \etal\ (1994), Kraft (1994), and Sneden (1998,1999).
Some general statements about cluster nucleosynthesis are summarized
here without attribution to specific papers, and the reader is strongly
encouraged to consult the reviews and the original papers 
quoted in them for details on these abundance trends.

{\it The CN cycle:} The chief products of ordinary CN cycle fusion
are observed at the surfaces of most RGB and AGB stars.
That is, the carbon isotope ratios are uniformly low
(4~$\leq$~\carbiso~$\leq$~10), carbon abundances are usually low
(--0.3~$\geq$~[C/Fe]~$\geq$~--1.3), and nitrogen abundances are correspondingly
very high (+0.5~$\leq$~[N/Fe]~$\leq$~+1.5).
However, the N overabundances are sometimes far greater than the amounts
that would be predicted from simple C$\rightarrow$N conversion.

{\it The ON cycle:} Globular cluster giants, unlike almost all halo
field giants, often exhibit very depleted oxygen abundances
(--1.0~$\leq$~[O/Fe]~$\leq$~+0.4).
This suggests that the ON cycle, which requires higher temperatures
($T$~$\sim$~40$\times$10$^6$~K) in hydrogen fusion zones than does the
CN cycle, has been active either in the giants that are being observed
or in an earlier cluster generation.
This cycle's major net effect is O$\rightarrow$N conversion, and can 
therefore account for the anomalously large N abundances mentioned above.   
Finally, in nearly all cluster giants with complete CNO abundance
data, the C+N+O abundance sum appears to be conserved, adding further
weight to the idea that the variations in these elements are simply
due to the combined CN and ON element re-shufflings.

{\it The NeNa cycle:} Sodium abundances also vary widely among globular
cluster giants (--0.3~$\leq$~[Na/Fe]~$\leq$~+0.4).
The same globular cluster giants that have low O abundances almost
invariably have high Na abundances; an anticorrelation between these
abundances apparently occurs in all lower metallicity clusters
([Fe/H]~$<$~--1) studied to date.
This anticorrelation suggests that the NeNa proton-fusion cycle, which
can work efficiently at the same temperatures as does the ON cycle,
has at some time in globular cluster histories converted Ne
(undetectable in cluster giant spectra) into Na.
 
{\it The MgAl cycle:} Aluminum abundances also have large star-to-star 
variations that are anticorrelated with O abundances, and in some 
well-studied clusters the anticorrelation extends also to Mg abundances.
Again, proton-capture fusion leading to Mg$\rightarrow$Al conversion
is the probable culprit (Shetrone 1996), but the burning temperature 
requirements (T~$\sim$~70$\times$10$^6$~K) are large enough that it 
is difficult to imagine low mass globular cluster giants performing 
the MgAl cycle, unless such such transmutations occur as the result 
of a thermal instability of the H or He shell source (Langer \etal\ 1997, 
Powell 1999).
Alternatively, stars with abnormally large Al abundances might either 
have been born with them, created in previous higher mass stars, or 
have accreted them from the winds of higher mass AGB stars.
 
{\it Other Nucleosynthesis effects:} Some significant cluster-to-cluster
abundance differences are seen in heavier elements that cannot be
altered in proton-capture synthesis reactions.
For example, the very heavy elements Ba, La, and
Eu can have very different abundance ratios in different clusters,
indicating varying contributions of slow and rapid neutron-capture
synthesis reactions to the creation of these elements.
Among the elements that participate in the major nuclear fusion
chains, silicon should only be altered during
the last stages of very high mass stars.
But its mean abundance varies from cluster to cluster; some globular
clusters have Si abundances nearly a factor of two larger than those
of typical halo field giants.
And in addition to the star-to-star variations of Al abundances within 
individual clusters, the Al mean abundance level also differs substantially 
from cluster to cluster.
All of these abundance anomalies point to nucleosynthesis contributions 
of multiple generations of stars in a given cluster, either from stars
that died before the present stars were born or during their formation.
Also, the relatively small numbers of high mass stars that must have 
existed in or preceded formation of each cluster probably produced 
supernovae of different masses in each cluster, creating distinct
``initial'' abundance distributions in each cluster.

\section{CN Bandstrengths in RGB and RGB Stars of NGC~6752}
 
Perhaps the first suggestion that AGB and RGB stars in some clusters
might on average have different compositions was made by Norris \etal\ (1981).
In a large-sample study of CN bandstrengths among giants of NGC~6752,
they found that there is a bi-modal distribution of
CN bandstrengths that is nearly independent of RGB position.
But they suggested that there is a nearly uni-modal set of CN bandstrengths
among the AGB stars: their CN bands are almost all weak.
Norris \etal\ presented this situation in their Figure~3, plotting
the CN absorption index S(3839) as a function of V magnitude and B--V colour.

We have used the formula developed by Norris \etal\ to convert S(3839)
to a CN bandstrength indicator that is independent of stellar
temperature/gravity effects, and in Figure~1 we show ``boxplots'' 
that illustrate the ranges in CN bandstrength found in RGB and AGB stars.
The lower CN strengths of the AGB stars on average is obvious,
but just as important is the near total lack of any CN strong AGB
stars in this cluster.  For comparison, we also show similar data for two
other clusters, M4 and M13.  The M4 CN bandstrength data are taken from 
either Norris (1981) or Suntzeff and Smith (1991), or the mean of both, 
where the variation of S(3839) with position in the c-m diagram has been 
removed according to Norris' formula.  The evolutionary status of the stars 
are those determined by Ivans \etal\, in their H-R diagram of 
Figure~12, which illustrates the reddening-free positions of the stars.  
For the M13 data, we referred to Suntzeff (1981), where we converted the 
photometric $m$(CN) indices to relative photometric bandstrengths 
$\delta$$m$(CN) using Suntzeff's suggested relationship of the lower 
limit of $m$(CN) to B--V colour index.  
We further transformed the $\delta$$m$(CN) values to 
$\delta$S(3839) relative bandstrengths employing the relationship we derived 
for $\delta$$m$(CN) and $\delta$S(3839) found using stars in common between 
the studies of NGC~6752 stars by Langer \etal\ (1992), who used $m$(CN), and 
Norris \etal\ (1981) who used S(3839).  The evolutionary status of the M13 
stars were those determined by Suntzeff (1981) and, in the cases where the 
photometry made the status ambiguous, we supplemented the information 
using the stars in common studied by Pilachowski \etal\ (1996b).  Thus,
the distributions shown in Figure~1 are all, in effect, on the 
$\delta$S(3839) system of Norris \etal\ (1981) and only include the stars
for which AGB vs RGB designations are unambiguous.

\begin{figure}
\epsfxsize=13cm 
\hspace{8.5cm}\epsfbox{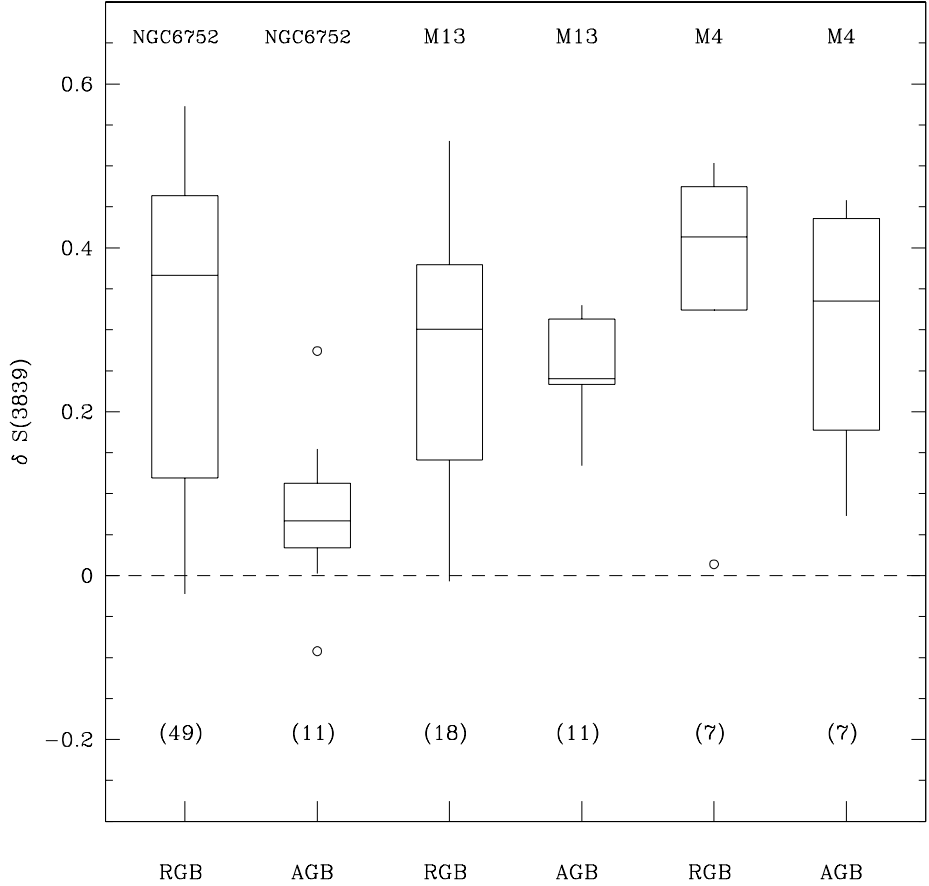} 
\caption[bhpt]{Simple boxplots illustrating CN bandstrength ranges in
RGB and AGB stars of NGC~6752, M13, and M4 are shown.  
For all of the individual abundance boxes, the horizontal line inside 
a box indicates the median value of $\delta$S(3839).
The vertical boundaries of a box show the interquartile range (the middle
50\% of the data).
The vertical tails extending from the boxes indicate the total range
of $\delta$S(3839), excluding outliers.
Mild outliers (those between 1.5 and 3 times the interquartile range) are
denoted by open circles.
No severe outliers (those greater than 3 times the interquartile range)
are present in these data.  The number of stars included in each boxplot is 
noted in parentheses.
The basic data (estimates of CN line blocking index S(3839)) were taken from
the following sources: 
(a) NGC6752: Norris \etal\ (1981), and we have used their suggested 
relationship of the lower limit of S(3839) to V magnitude to compute the 
relative CN bandstrengths $\delta$S(3839);
(b) M13: converted photometric CN index $m$(CN) values obtained by
Suntzeff (1981).  We first removed the variation of the index due to 
position in the c-m diagram using Suntzeff's formula, and then converted
the results to $\delta$S(3839) values using the relationship we found for the 
NGC~6752 stars in common to the $m$(CN) values presented by Langer \etal\ 
(1992) and the $\delta$S(3839) values of Norris \etal\ (1981).  
(c) M4: Norris (1981) or Suntzeff \& Smith (1991), or the mean of both, 
using Norris' suggested relationship of the lower limit of S(3839) to V 
magnitude.}
\end{figure}

Norris \etal\ (1981) offered two possible explanations for the relatively
weak CN bandstrengths in NGC~6752 AGB stars; both explanations
involve an inability of the strong-CN RGB stars to ascend the giant 
branch a second time after HB evolution.	
In one scenario, some cluster stars would have been born with abnormally 
large C and/or N abundances, accompanied by larger-than-average He/H 
ratios (presumably from the CN and/or ON cycles).
Stellar evolution computations (\eg, Lee \etal\ 1994, and 
references therein) have shown that RGB stars with higher He 
contents will, after they undergo the He flash, take up residence in
bluer parts of the HB than do otherwise identical stars with lower
He contents.
In fact, these stars may arrive at such a blue HB position that they
may eventually evolve directly to the white dwarf track, entirely
avoiding the AGB stage.
In the other scenario, larger internal mixing in CN-strong stars 
during RGB evolution might drive large amounts of mass loss, leading 
to lower-than-average envelope masses after the He flash.
Again, such stars would wind up on the bluer end of the HB, possibly
never to return as AGB stars.
Thus the stars observed on the AGB of NGC~6752 may have weak CN bands
because they are the former RGB stars that had little mixing of CN
cycle products (N and He) into their envelopes;
they are the ones that survived the HB stage to rise again toward
the giant branch tip.
 
This latter hypothesis can be tested by comparing abundances of
light proton-capture elements (C, N, O, Na, Mg, Al) in AGB
and RGB stars of NGC~6752.
Unfortunately, the extant high resolution spectroscopic studies
of NGC~6752 giants (Gratton 1987, Norris \& Da Costa 1995, 
Minniti \etal\ 1996, Shetrone 1998) were only able to include 
the brightest stars near the RGB tip, where the distinction between
RGB and AGB stars cannot be made.

\section{Sodium Abundance Variations in M13 Giants}

Pilachowski \etal\ (1996b) derived sodium abundances for 130 giants in M13;
their program stars ranged from those at the RGB tip to ones about
as faint as the HB.
They showed that Na abundances of most M13 giants are greater than those of
similar-metallicity halo field stars, but there are some significant
differences between RGB and AGB stars in this cluster.
We illustrate this situation in Figure~2 with another boxplot,
in which we compare [Na/Fe] ratios for lower luminosity M13 RGB stars
(those with log~$g$~$>$~1), RGB tip stars (log~$g$~$<$~1), and AGB
stars, along with [Na/Fe] ratios for field stars in the metallicity
range --1.2~$>$~[Fe/H]~$>$~--1.9 (Pilachowski \etal\ 1996a).
The higher Na abundances of M13 giants is obvious in this figure,
but it is also clear that the AGB stars have lower mean Na abundances
than do the RGB tip stars, and that they have a narrower range in Na.
It is possible that the Na abundances for the very cool RGB tip stars
must be corrected downward somewhat to correct for departures from LTE
(\eg, Gratton \etal\ 1999). 
But oxygen abundances in cluster giants are always determined from
the [O~I] transitions, which do not suffer substantial departures from LTE.
And in M13 not only do the RGB stars exhibit on average the largest 
Na abundances but they also have the lowest O abundances (Kraft \etal\
1997).
Therefore the difference between the the mean levels of Na in AGB and
RB stars in M13 is probably real.
 
\begin{figure}
\epsfxsize=13cm 
\hspace{18.5cm}\epsfbox{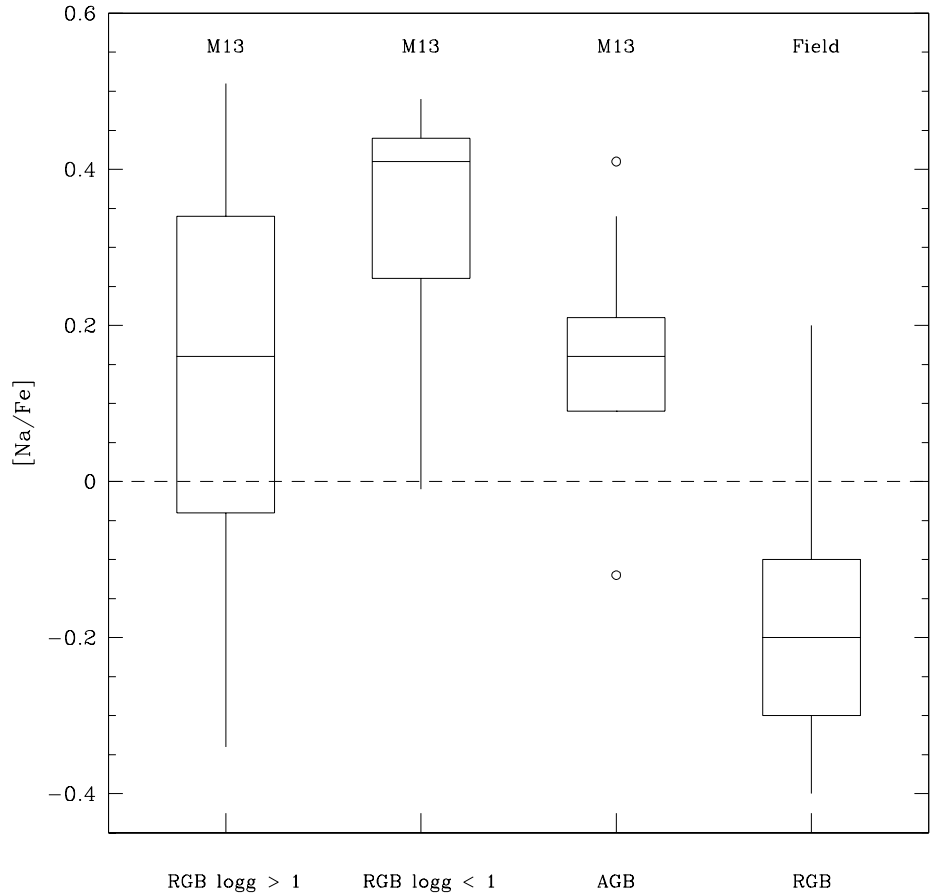} 
\caption[bhpt]{Boxplots of sodium abundances in M13 (Pilachowski \etal\
1996b) and field giant stars (Pilachowski \etal\ 1996a).
The statistical abundance distributions represented by each box's
vertical boundaries, etc., are as described in Figure~1.
The different boxes illustrate the sodium abundance distributions      
found for RGB tip stars, lower luminosity giants,
and AGB stars.
For contrast, a boxplot representing sodium abundances in field giants
is also shown in this figure.}
\end{figure}

Pilachowski \etal\ (1996) followed a line of reasoning similar to that
of Norris \etal\ (1981) in supposing that the presently observed AGB 
stars in M13 are those whose envelope He contents remained relatively 
low when they were RGB stars; the RGB stars with elevated He took up 
residence on the blue part of the HB and never arrived on the AGB. 
However, the Pilachowski \etal\ scenario differed from that of Norris 
\etal\ in one important respect: the RGB stars with elevated He were 
those that had contaminated their atmospheres with material that had 
been processed through the CNO hydrogen-burning shell, in accordance 
with the deep mixing scenario and nuclear transmutation calculations 
of Langer \etal\ (1993), Langer \& Hoffman (1995) and 
Cavallo \etal\ (1998). 
Sweigart (1997a,b) showed that such ``deep mixed'' stars could 
indeed be moved sharply to the blue in their subsequent evolution
onto the HB, largely as a result of increased mass loss prior 
to the helium core flash. 
Pilachowski \etal\ also noted that since M13 has the most extreme
cases of Na and Al enhancements and O depletions among RGB stars, 
it also probably has a higher percentage of high-He stars than other 
globular clusters.  
If so, then the AGB of M13 ought to be relatively unpopulated. 
This view is supported by the statistics of Caputo \etal\ (1978) and 
Buzzoni \etal\ (1983), from which one finds that M13 has the lowest 
ratio (by a factor of $\sim$2) of AGB to RGB stars among the 16 
clusters studied.

\section{Some Additional Comments}
 
M13 and NGC~6752 represent the clearest cases for chemical composition
differences between AGB and RGB stars in globular clusters.
But truth in advertising compels us to admit that the situation
is probably far more complex than we have suggested so far.
Smith \& Norris (1993) suggested that the AGB stars of M5 have
a different CN bandstrength distribution: ``... the observations 
reported in this paper yield no consistent picture of the CN distributions
among stars in more advanced stages of evolution.  The asymptotic giant
branch appears to be deficient in CN-weak stars for M5, but 
deficient in CN-strong stars for NGC~6752.''

Consideration of these differences has been made possible by the
existence of very large bandstrength or abundance samples in these
two clusters.
Unfortunately, most other globulars have not been studied in sufficient
detail to assess the chemical compositions of AGB stars.
In their study of a large number of bright giants in M4, Ivans \etal\ 
(1999) found some of the same correlated variations in proton-capture 
elements that have been seen in other clusters.
Their data were most extensive for the determination of oxygen abundances,
and they concluded that the mean oxygen abundance of M4 AGB stars is
slightly larger than that of the RGB stars.
This provides mild further support for the suggestion that AGB stars
in globular cluster are on average less chemical evolved in the
proton-capture elements than are RGB stars.
 
This problem cannot be effectively dealt with until stellar samples
in many globular clusters include at least 10 AGB stars, as well
as many more RGB stars over a large luminosity range.
Enough detailed high resolution, large wavelength coverage spectroscopic
studies of individual stars in selected globulars to make it
clear that the proton-capture phenomenon is ``universal''.
Thus in addition to the continued full-scale abundance analyses
of the brightest cluster members, it will be especially fruitful
to now survey cluster giant branches with multi-object spectrometers 
(in the manner of Pilachowski \etal\ 1996b) that concentrate on fairly 
complete descriptions of the abundance trends of just one or two elements
that will stand as surrogates for the behaviour of the whole set of
proton-capture elements.

\acknowledgements
We thank Raffaele Gratton for helpful discussions on this work.
This research was supported by NSF grants AST-9217970 to RPK
AST-9618364 to CS.
Travel support given by the Rome Observatory to CS is gratefully acknowledged.

\end{document}